\documentstyle{article}
\title{Stochastic quantum field dynamics
in the proper time}
\author{Z. Haba\\Institute of Theoretical Physics,University of
Wroclaw,Wroclaw,Poland}
\date{}
\begin{document}
\maketitle
\begin{abstract}
We consider a quantization of relativistic wave equations
which allows to treat quantum fields together with
interacting particles at a finite time.
We discuss also a dissipative interaction with the environment.
We introduce a stochastic wave function whose dynamics is
determined by a non-linear Schr\"odinger-type evolution
equation in an additional time parameter.
 The correct classical limit
requires  the proper time interpretation of the time parameter.
An average over the proper time leads to the conventional
quantum field theory of particles which are free at an infinite
space separation. We consider models with scalar and vector
fields on a pseudoriemannian manifold.
A quantization of the Einstein gravity in this approach is briefly
discussed.

\end{abstract}
{\bf PACS NUMBERS}: 03.70.+k,02.50.Ga,04.60.-m
\pagestyle{myheadings}
\markboth{Stochastic quantum field}{Stochastic quantum field}
\renewcommand{\thesection}{\Roman{section}}
\section{Introduction}
The classical non-relativistic mechanics describes system$^{\prime}$s history
${\bf x}_{t}$ as a function of time ( a succession of events).
In a classical relativistic theory time and space
 should be treated on an equal footing.
If so then we should  consider world lines (histories) $x_{\mu}(\cdot)$
rather than events. In quantum mechanics the wave function $\psi_{t}({\bf x} )$
gives an amplitude of probability of detecting a particle at
the position ${\bf x}$. Only in the classical limit
 $|\psi_{t}({\bf x})|^{2}
\approx |\psi({\bf x}_{t})|^{2} $ we regain the time-dependent trajectory.
By an analogy to the non-relativistic quantum mechanics
in  an explicitly relativistic treatment we would require
that in the classical limit the  probability of an occurence of the history
$x_{\mu}(\cdot)$ in a state $\psi$ should be
$|\psi(x_{\mu}(\cdot))|^{2}$. This probability amplitude can vary
 depending on a certain universal time $\tau$ which should be
 a Lorentz scalar. In  classical relativistic physics
 there is a good candidate for this universal time (the proper time)
 which describes e.g. the periods of oscillations of a monochromatic light
 emitted by an atom or the frequencies
  of crystal vibrations in their rest frames.

The relativistic quantum mechanics is usually rejected in favor of
the quantum field theory (QFT). There is at least one good reason for that:
the need to describe the processes of particle creation and annihilation.
The QFT successfully and consistently treats the scattering processes
with particle creation and annihilation. However, in order to describe
bound states (e.g. the Lamb shift in the hydrogen atom ) we resign
of a relativistic description and combine quantum field theory
with  relativistic quantum mechanics. It seems that the conventional
QFT is unable to treat any space-time description of relativistic
particles. In particular, the limit $\hbar\rightarrow 0$
of quantum field theory exists \cite{hepp} but it describes
the classical field theory rather than the classical mechanics.
  Nevertheless, we would still be interested e.g.in the
probability density of finding a relativistic electron
close to the nucleus and a classical limit of this probability.

In this paper we consider a non-linear relativistic quantum wave mechanics
which is inseparably bound with the quantum field theory.
The quantum effects are achieved by an addition of a noise
to the relativistic field wave equation. It is shown that the
expectation values of the random part  of the field
after an average over the proper time coincide with the
time-ordered vacuum expectation values of the conventional
QFT.

\section{Relativistic wave equation}
We assume that the evolution of the wave function $\psi_{\tau}$ is
determined by its initial value $\psi$. In a classical limit
 we can imagine $\psi$ as
 a wave packet $\psi (x)$ concentrated on a certain
 space-time point.
We consider a relativistic particle in an electromagnetic field A.
We suggest the following
 analogue of the Schr\"odinger equation
(see earlier papers on such wave equations \cite{stu}\cite{hor}\cite{feyn}\cite{col})
\begin{equation}
i\hbar\partial_{\tau}\psi=\frac{1}{2M}g^{\mu\nu}
(-i\hbar\partial_{\mu} +\frac{1}{c}A_{\mu})(-i\hbar\partial_{\nu}+
\frac{1}{c}A_{\nu})\psi\equiv
-\frac{\hbar^{2}}{2M}\Box_{A}\psi
\end{equation}
where c is the velocity of light and $g^{\mu\nu}=(1,1,1,-1)$ is the Minkowski metric ( we
shall also use the notation $R^{d}$ for the Minkowski space suggesting
an arbitrary dimension d of this space).
If we write  ( where $W_{\tau}$ may be complex )
\begin{equation}
\psi_{\tau}=exp(\frac{i}{\hbar}W_{\tau})
\end{equation}
Then, from eq.(1) it follows
\begin{equation}
\partial_{\tau}W_{\tau}+\frac{1}{2M}(\partial_{\mu}W_{\tau}+\frac{1}{c}A_{\mu})
(\partial^{\mu}W_{\tau}+\frac{1}{c}A^{\mu})
 -\frac{i\hbar}{2M}(\Box W_{\tau} +\frac{1}{c}\partial_{\mu}A^{\mu})=0
\end{equation}
In a formal limit $\hbar\rightarrow 0$ we obtain the Hamilton-Jacobi equation
\begin{equation}
\partial_{\tau}W_{\tau}+\frac{1}{2M}(\partial_{\mu}W_{\tau}+\frac{1}{c}A_{\mu})
(\partial^{\mu}W_{\tau}+\frac{1}{c}A^{\mu})=0
\end{equation}
Eq.(4) does not coincide with the conventional Hamilton-Jacobi
equation in classical  relativistic dynamics \cite{landautp}\cite{goldstein}
which has no $\partial_{\tau}W$ term. Nevertheless, eq.(4)
can be derived from classical mechanics. Let us recall that if the
relativistic Lagrangian  is chosen in the form invariant
under the reparametrization $x(\gamma)\rightarrow x(f(\gamma))$
then the canonical Hamiltonian
\begin{equation}
H=\frac{1}{2M}(p_{\mu}+\frac{1}{c}A_{\mu})(p^{\mu}+\frac{1}{c}A^{\mu})
\end{equation}
is identically equal to zero. This constraint H=0
\cite{dirac} generates correct equations of motion if the time
parameter is interpreted as the proper time.
If from the beginning we choose $\gamma$ as the proper time
then $H\neq 0$. In such a case we may
pose the problem of a canonical change of coordinates
( determined by the generating function $W$ )
such that in the new coordinates $H\rightarrow H+\partial_{\tau}W=0$.
The generating function W (hence also the classical dynamics)
 is defined by the solution of
 the Hamilton-Jacobi equation (4).
 Eq.(1) can be considered as a quantization of eq.(4).
 In the standard quantization scheme
of constrained systems \cite{dirac} one argues
that the quantum theory should be invariant under the
choice of the parameter $\gamma$   ( reparametrization invariance).
Hence, the dependence on this parameter is gauged away and
what remains is the Klein-Gordon equation $H\psi= 0 $.
However,  in our interpretation the proper time has a physical meaning.
Hence, we make this preferred choice of $\gamma$ in the Lagrangian.
In such a case the canonical Hamiltonian (5) is different from zero.
It again generates the correct equations of motion. Then, the conventional
quantiztion scheme leads to eq.(1) rather than to the Klein-Gordon equation.
We show below that conversely the classical dynamics (4) is determined
as a limit $\hbar\rightarrow 0$ of the quantum dynamics (1) if
$\tau$ is identified with the classical proper time.

For further purposes we write the metric tensor in eq.(1) in terms of
complex  vierbeins $v$
\begin{equation}
ig^{\mu\nu}=v^{\mu}_{a} v^{\nu}_{a}
\end{equation}
We can take as a solution of eq.(6)
\begin{displaymath}
v^{k}_{a}=\lambda\delta^{k}_{a}
\end{displaymath}
if k=1,2,3 and
\begin{displaymath}
v^{0}_{a}=\overline{\lambda}\delta^{0}_{a}
\end{displaymath}
where
\begin{displaymath}
\lambda=\sqrt{i}=\frac{1}{\sqrt{2}}(1+i)
\end{displaymath}
The Feynman integral \cite{feyn}
 supplies a simple intuitive way of
proving the classical limit.
We assume that all functions we
deal with are analytic. Then, we can
express the solution of eq.(1) by the following rigorous form of the
 Feynman integral
( $\tau \geq 0$, see refs. \cite{haba}\cite{haba3} for a more  precise formulation and
all assumptions )

\begin{equation}
\begin{array}{l}
\psi_{\tau}(x)=E\Big[exp\Big(\frac{i}{c\hbar}\int_{0}^{\tau}
ds A_{\mu}(x+\sigma v b_{s})\sigma v^{\mu}_{a}db^{a}_{s}\Big)
\psi\big(x+\sigma v b_{\tau})\Big]
\end{array}
\end{equation}
where $b$ is the Brownian motion i.e. the real Gaussian process
with values in $R^{d}$ and  the covariance
\begin{displaymath}
E[b^{a}(s)b^{c}(\tau)]=\delta^{ac}min(s,\tau)
\end{displaymath}
In eq.(7)
\begin{displaymath}
\sigma=\sqrt{\frac{\hbar}{M}}
\end{displaymath}

We write eq.(1) and eq.(7) still in another (equivalent) form.
Let $ W_{\tau}$ be the solution of the Hamilton-Jacobi equation (4) with
the initial condition W . We express the wave function $\psi $
as a product
\begin{equation}
\psi_{\tau}=exp(\frac{i}{\hbar}W_{\tau})\Phi_{\tau}
\end{equation}
If $\psi_{\tau}$ is a solution of eq.(1) then $\Phi_{\tau}$ fulfills the equation   ( assuming  the Lorentz
gauge $  \partial_{\mu}A^{\mu} =0$ )
\begin{equation}
\partial_{\tau}\Phi_{\tau}\equiv -\frac{i}{\hbar}\tilde{H}\Phi_{\tau}=
\frac{i\hbar}{2M}\Box \Phi_{\tau}
-\frac{1}{M}(\partial^{\mu} W_{\tau}+ \frac{1}{c}A_{\tau}^{\mu})
\partial_{\mu}\Phi_{\tau}
-\frac{1}{2M}\Box W_{\tau} \Phi_{\tau}
\end{equation}
where
\begin{displaymath}
\Box=g^{\mu\nu}\partial_{\mu}\partial_{\nu}
\end{displaymath}
The equivalent form of the Feynman formula follows from eq.(9).
So, if $q_{\mu}(\tau) $ is the solution of the equation
( $0\leq s\leq  \tau$ )
\begin{equation}
dq_{\mu}=-\frac{1}{M}\Big(\partial_{\mu}W(\tau-s,q(s))+
\frac{1}{c}A_{\mu}(q(s)) \Big)ds
+\sigma v_{\mu a}db^{a}(s)
\end{equation}
where $W(\tau)$ is the solution of the Hamilton-Jacobi  equation (4), then the
solution of eq.(1) with the initial condition $\psi=exp(\frac{i}{\hbar}W)\Phi
$  reads

\begin{equation}
\begin{array}{l}
\psi_{\tau}( x)=exp\Big(\frac{i}{\hbar}W(\tau, x)\Big)
E\Big[ exp\Big(-\int_{0}^{\tau}\frac{1}{2M}\Box W(\tau-s,q(s))ds\Big)
\Phi(q(\tau))\Big]
\end{array}
\end{equation}
In the limit $\hbar\rightarrow 0$ of the stochastic process we obtain
the flow  (here $0\leq s\leq \tau$)
\begin{equation}
\frac{d{\bf \xi}_{\mu}}{ds}=-\frac{1}{M}\big(\partial_{\mu}
 W(\tau-s,\xi(s))+\frac{1}{c}A_{\mu}(\xi(s))\big)
\end{equation}
Till $O(\hbar) $ terms we have in the semi-classical approximation

\begin{equation}
\begin{array}{l}
\psi_{\tau}(x)^{cl}=exp\Big(\frac{i}{\hbar}W(\tau,x)\Big)
\Phi\big(\xi(\tau,x)\big)
 exp\Bigg(-\int_{0}^{\tau}\frac{1}{2M}\Box W(\tau-s,\xi(s))ds\Bigg)
\end{array}
\end{equation}
If we differentiate eq.(12) once more over s and make use
of the Hamilton-Jacobi equation (4) then we conclude
that $\xi$ fulfills the equation
\begin{equation}
M\frac{d^{2}\xi_{\mu}}{ds^{2}}=F_{\mu\nu}(\xi)\frac{d\xi^{\nu}}{ds}
\end{equation}
where
\begin{displaymath}
F_{\mu\nu}=\partial_{\mu}A_{\nu}-\partial_{\nu}A_{\mu}
\end{displaymath}
with the boundary conditions
\begin{displaymath}
\xi(s)_{|s=0}=x
\end{displaymath}
\begin{equation}
\frac{d\xi_{\mu}}{ds}_{|s=\tau}=-\frac{1}{M}(\partial_{\mu}W(x)+
\frac{1}{c}A_{\mu}(x))
\end{equation}
We could also obtain eq.(14) directly from the Feynman formula (7).
For this purpose we make a shift of variables $b^{a}\rightarrow
b^{a}+f^{a}$. Then, for any functional $\chi$   of the Brownian motion
( $b(s), s\leq\tau$ ) we have ( the Cameron Martin formula \cite{ikeda})
\begin{equation}
E[\chi(b)]=E\Big[exp\Big(-\int_{0}^{\tau}f^{a}db^{a}
-  \frac{1}{2} \int_{0}^{\tau}\frac{df^{a}}{ds}\frac{df^{a}}{ds} ds\Big)
\chi(b+f)\Big]
\end{equation}
We apply the shift (16) to the Feynman  formula (7) with
$\psi=exp(\frac{i}{\hbar}W)\Phi$ and   $\sigma vf=\xi-x$ ,  where
$\xi$ is the solution of the equation   (14) with the boundary conditions (15).
Then, expanding in $\sigma$ we obtain the formula (13) till $O(\sigma)$.

Summarizing the result (14) derived either from the Feynman formulas (7)
and (11)
or directly from the Schr\"odinger equation
in the form (9) we can say that the limit
$\hbar\rightarrow 0$ exists if and only if
there exists $\xi$ such that the equations of motion (14) are
satisfied. As an immediate but important consequence
 of the equations of motion (14) we obtain
\begin{displaymath}
\frac{d^{2}\xi_{\mu}}{ds^{2}}\frac{d\xi^{\mu}}{ds}=0
\end{displaymath}
Hence, the square of the covariant velocity is time-independent.
From the boundary condition (15) we obtain that
\begin{displaymath}
\frac{d\xi_{\mu}}{ds}\frac{d\xi^{\mu}}{ds}=\frac{1}{M^{2}}(\partial_{\mu}W(x)
+\frac{1}{c}A_{\mu}(x) )
     (\partial^{\mu}W(x)+\frac{1}{c}A^{\mu}(x) )
 \end{displaymath}
 If in the Hamilton-Jacobi equation (4)
 \begin{equation}
 \partial_{\tau}W_{\tau} =\frac{Mc^{2}}{2}
 \end{equation}
 Then
\begin{equation}
\frac{d\xi_{\mu}}{ds}\frac{d\xi^{\mu}}{ds}=-c^{2}
 \end{equation}
 Eq.(18) means that $s$ ( hence also $\tau$ in the Schr\"odinger equation)
 is the proper time.
 Moreover, this interpretation must be preserved in higher orders
  of $\hbar$ because the subsequent quasiclassical expansion
  is determined by the first order. It is well-known
 from classical relativistic dynamics that the equations of
 motion (14) with a time parameter s have a correct form only if
 s coincides with the proper time. If the dependence of $W_{\tau}$
 on $\tau$ in eq.(8) is determined by eq.(17) then
 $exp(\frac{i}{\hbar}W_{\tau}(x))$ also solves  the Klein-Gordon
 equation ( in such a case we regain the conventional relativistic wave
 equation ) . On the other hand $\psi_{\tau}\approx
 exp(\frac{i\tau Mc^{2}}{2\hbar} )$ means that the wave function
 oscillates rapidly with a very small period
 of oscillations .
 It is known from classical mechanics \cite{landau} that
 such additional rapid oscillations have little effect on the mean
 behavior in a slowly varying potential. We suggest that  even if
 the condition (17) is not satisfied then the evolution in the proper
 time describes rapid oscillations which have little effect
 on the evolution in the Minkowski time
 for most physically relevant potentials $A_{\mu}$. Hence, after an averaging
 over the proper time we obtain the conventional quantum theory.
            
For an understanding of eq.(1)  it is important to see its non-relativistic
limit ( the problem is discussed in a different way in ref. \cite{horob}  )
. The non-relativistic energy $\epsilon$ is related to the
relativistic energy $p_{0}$ , mass M and the momentum ${\bf p}$ by the formula
\begin{displaymath}
\epsilon=c({\bf p}^{2}+M^{2}c^{2})^{\frac{1}{2}}-Mc^{2}=
\frac{{\bf p}^{2}}{2M}+o(\frac{1}{c})
\end{displaymath}
Let
\begin{displaymath}
\psi_{\tau}=exp(\frac{-iMc^{2}(\tau-2t)}{2\hbar})\tilde{\psi_{\tau}}
\end{displaymath}
Then, in the limit $c\rightarrow \infty$ we obtain  ( we write
$A=({\bf A},V)$ i.e.$A_{0}=V$)
\begin{equation}
i\hbar\partial_{\tau}\tilde{\psi_{\tau}}=-i\hbar\partial_{t}\tilde{\psi_{\tau}}
+\frac{1}{2M}(-i\hbar\nabla +\frac{1}{c}{\bf A})^{2}\tilde{\psi_{\tau}}+V({\bf x},t)
\tilde{\psi_{\tau}} \equiv K\tilde{\psi_{\tau}}
\end{equation}
Strictly speaking we should have omitted the $\frac{1}{c}{\bf A}$ term
in the limit $c\rightarrow \infty$ but we keep it in order
to be in agreement with the conventional procedure.
If the potentials A and V are $t$-independent
then   we can express eq.(19)
as the Schr\"odinger equation with a new time $\hat{t}=t+\tau$. In such a case
$\tau $ is just a global shift of time in the non-relativistic
quantum mechanics.

If the potentials are time-dependent then we  obtain
 Howland$^{\prime}$s description \cite{howland} of
 the time evolution in
time-dependent potentials .
In such a case t is treated as a coordinate on an equal
footing with ${\bf x}$ ( the eigenvalues of K are called quasienergies
and constitute a standard tool in an investigation of time-dependent systems).

The relation between the $\tau$-evolution and the conventional Schr\"odinger
evolution $U_{sch}$ is well-known \cite{howland}
\begin{equation}
exp(-\frac{i\tau}{\hbar}K)\tilde{\psi}(t,{\bf x})=
U_{sch}(t,t-\tau)\tilde{\psi}(t-\tau,{\bf x})
\end{equation}
In particular, it follows that the scattering theories in terms
of $exp(-\frac{i\tau}{\hbar}K)$ and $U_{sch}$ are equivalent.

\section{Quantum free fields and free particles}

Let us consider a time evolution of the  wave function
with no interaction
\begin{equation}
i\hbar\partial_{\tau}\phi_{\tau} =-\frac{\hbar^{2}}{2M}\Box\phi_{\tau}
\end{equation}
An easy computation shows that the wave packet
\begin{displaymath}
\phi(x)=\int dp \tilde{\phi}(p)exp(\frac{i}{\hbar}p_{\mu}x^{\mu})
\end{displaymath}
evolves into
\begin{equation}
\phi_{\tau}(x)=\int  dp \tilde{\phi}(p)  exp(-\frac{ip^{2}\tau}{2M\hbar})
exp(\frac{i}{\hbar}p_{\mu}x^{\mu})
\end{equation}
If $\tilde{\phi}(p) $ is regular in $\hbar$ and its support is concentrated at
$p_{c}$ e.g.
 \begin{displaymath}
 \tilde{\phi}(p)=exp\Big(-\frac{1}{2\mu}({\bf p}-{\bf p}_{c})^{2}
-\frac{1}{2\mu}( p_{0}+ p_{0c})^{2}\Big)
 \end{displaymath}
 then  we can conclude that
\begin{equation}
\phi_{\tau}(x)\approx \phi(x-\frac{p_{c}\tau}{M})
\end{equation}
Eq.(23) correctly describes the evolution of a relativistic wave packet
if $\tau$ is interpreted as the proper time.

We can describe a wave packet of k-particles by a generalization of eq.(22)
\begin{displaymath}
\phi(x(1),....,x(k))=\int dp(1)....dp(k)
 \tilde{\phi}(p(1),....,p(k))exp\Big(\sum_{j=1}^{k}
 \frac{i}{\hbar}p_{\mu}(j)x^{\mu}(j)  \Big)
\end{displaymath}
Its time evolution is determined by a generalization of eq.(21)
\begin{equation}
i\hbar\partial_{\tau}\phi(x(1),....,x(k))=-\frac{\hbar^{2}}{2M}
\sum_{j=1}^{k}\Box_{j}\phi(x(1),....,x(k))
\end{equation}
The solution of eq.(24) shows an independent  free evolution
of each particle
\begin{displaymath}
x_{\tau}(j)=x(j)+\frac{p_{c}(j)\tau}{M}
\end{displaymath}
By  a free quantum field we understand
an object which can describe any number of free particles
as excitations of the field at any point
of the space-time.  By its physical meaning the quantum field
of a large intensity is supposed
to behave as its classical  counterpart ( keep in mind the example of
 the electromagnetic
field ) . It should also describe  fluctuations corresponding to
the Heisenberg  uncertainty principle. We represent these fluctuations
by the Gaussian field B with  the covariance
\begin{equation}
E[B(\tau,x)B(s,y)]=min(\tau,s) \delta(x-y)
\end{equation}
We quantize the wave equation (21) by adding the noise
( when $M\rightarrow \infty $ then $\phi $ is the Brownian motion)
\begin{equation}
d\phi(\tau,x)=\frac{i\hbar}{2M}\Box\phi(\tau,x)d\tau+\sqrt{2}\hbar dB(\tau,x)
\end{equation}
The solution of  eq.(26) is a sum
of two pieces ( see the subsequent section):
the first part is the wave function (21) and the second one is a noise.
 We describe
the quantum field by means of the random part  relating
the correlation functions of the
random field to time-ordered products of non-commutative
operators.

The solution of eq. (26)  depends on the proper
time parameter $\tau$.
We suggest that  correlations which are observed in experiments
result from an average over the rapid oscillations in the proper time.
We shall show that
time-ordered vacuum expectation values
of the conventional QFT of massless particles coincide with an average over the proper time
\begin{equation}
<\phi(x_{1})....\phi(x_{k})>=lim_{T\rightarrow\infty}
(T-\tau_{0})^{-1}\int_{\tau_{0}}^{T} d\tau E[\phi(\tau,x_{1})
.....\phi(\tau,x_{k})]
\end{equation}
For massive particles with mass M we should add the term
 $\frac{\hbar}{2i}M\phi$ on the r.h.s. of eq.(26) i.e.
 $\frac{1}{2M}\Box\rightarrow \frac{1}{2M}(\Box -M^{2})$
 (in eq.(1) M fulfills the role of the mass only in the non-relativistic
 limit).

There is an alternative to the stochastic quantization (26) of
the wave equation (21). We could simply treat eq.(21) as an operator equation
in the Fock space. Then , the conventional quantum free field
is a particular $\tau$-independent solution of eq.(21). We could add an
interaction and continue the proper time quantization in the operator formalism.
We outline such an approach in the Appendix.

\section{Mathematical aspects of the linear stochastic equation}
We consider a general form of the linear stochastic differential
equation
\begin{equation}
d\phi_{\tau}=-i{\cal A}\phi_{\tau}d\tau +\sqrt{2}\hbar dB_{\tau}
\end{equation}
Eq.(28)  is treated as an equation in the Gelfand triple \cite{gelfand}
\begin{displaymath}
{\cal S}(R^{d})\subset L^{2}(R^{d})\subset {\cal S}^{\prime}(R^{d})
\end{displaymath}
where ${\cal S}^{\prime}(R^{d})$ is the Schwartz  space
of tempered  distributions.
${\cal A}$ is a real self-adjoint operator in $L^{2}(R^{d})$.
Hence, there exists  the unitary group U
\begin{equation}
U(\tau)=exp(-i\tau{\cal A}) =cos(\tau {\cal A})-i\sin(\tau{\cal A})
\end{equation}
The solution of eq.(28) with the initial condition $\phi$ at $\tau_{0}$
reads
\begin{equation}
\phi_{\tau}= U(\tau-\tau_{0})\phi+\sqrt{2}\hbar \int_{\tau_{0}}^{\tau}
 U(\tau-s)dB_{s}
\end{equation}
It is common to work with real processes. For this purpose let us
decompose the complex field $\phi=\phi_{1}+i\phi_{2}$ into
its real and imaginary parts
\begin{equation}
\phi_{1}(\tau)=cos({\cal A}(\tau-\tau_{0}))\phi_{1}  +
sin({\cal A}(\tau-\tau_{0}))\phi_{2}+\sqrt{2} \hbar
\int_{\tau_{0}}^{\tau}cos({\cal A}(\tau-s))dB_{s}
\end{equation}
and

\begin{equation}
\phi_{2}(\tau)=cos({\cal A}(\tau-\tau_{0}))\phi_{2}  -
sin({\cal A}(\tau-\tau_{0}))\phi_{1}-\sqrt{2} \hbar
\int_{\tau_{0}}^{\tau}sin({\cal A}(\tau-s))dB_{s}
\end{equation}
The solutions $\phi_{1}(\tau)$ and $\phi_{2}(\tau)$ determine the
transition function ( from the initial point $(\phi_{1},\phi_{2})$ to
a set $\Gamma\subset{\cal S}^{\prime}$ ) of the stochastic process
\begin{equation}
P(\tau_{0},\phi_{1},\phi_{2};\tau,\Gamma)=
P((\phi_{1}(\tau),\phi_{2}(\tau))\in \Gamma)
\end{equation}
The stochastic process $\phi_{\tau}$ determines a solution of
the differential
equation
\begin{equation}
\partial_{\tau}\Phi_{\tau}=\hbar^{2}Tr(D^{2}\Phi_{\tau})-i({\cal A}\phi, D\Phi_{\tau} )
\end{equation}
where
\begin{displaymath}
(D\Phi(\phi) ,f)=lim_{\epsilon\rightarrow 0}\epsilon^{-1}
\Big(\Phi(\phi+\epsilon f) -\Phi(\phi)\Big)
\end{displaymath}
is the Frechet derivative ; the second order derivative
$D^{2}\Phi$ is an operator whose trace gives the Laplacian in an infinite
number of dimensions ( for a diffusion in infinite dimensional spaces
see \cite{daletsky}).   In physicists$^{\prime}$ notation
\begin{displaymath}
({\cal A}\phi, D\Phi)=\int dx ({\cal A}\phi)(x)
 \frac{\delta \Phi}{\delta \phi(x)}
\end{displaymath}
and
\begin{displaymath}
Tr(D^{2}\Phi)=\int dx
 \frac{\delta^{2} \Phi}{\delta \phi(x)\delta \phi(x)}
\end{displaymath}
The solution of the differential equation (34) with the initial condition
$\Phi$ reads
\begin{equation}
\Phi_{\tau}(\phi)=E\Big[\Phi\Big(\phi_{\tau}(\phi)\Big)\Big]
\end{equation}
where $ \phi_{\tau}(\phi) $ is the solution of eq.(28) with the real initial
condition $\phi$ at $\tau_{0}$.

We can also express the solution by means of the transition probability (33)
\begin{equation}
\Phi_{\tau}(\phi)=\int P(\tau_{0},\phi,0;\tau,d\phi_{1}^{\prime},d\phi_{2}^{\prime})
\Phi(\phi_{1}^{\prime}+i\phi_{2}^{\prime})
\end{equation}
$P (\tau,\phi,\Gamma)$ also solves eq.(34)
with a $\delta$-type initial condition . The transition probability P is
defined by the Gaussian measure
which is uniquely determined by its mean and covariance.
The mean and the covariance can be calculated
explicitly from eqs.(31)-(32).
It can be seen from these equations that the average (27) over
$\tau$ does not exist separately for $\phi_{1}$ and $\phi_{2}$.
We show however that the average (27) does exist for the complex field
$\phi=\phi_{1}+i\phi_{2}$. First of all, concerning the mean value
\begin{equation}
\int_{\tau_{0}}^{T}d\tau (f,exp(-i{\cal A}(\tau-\tau_{0}))\phi)=
i((exp(-i(T-\tau_{0}){\cal A})-1){\cal A}^{-1}f,\phi )
\end{equation}
Hence, the limit (27) of the mean value is equal
to zero if ${\cal A}$ is invertible on f i.e. if  $f\in Range({\cal A})$.
On the other hand if we consider ${\cal A}=-\frac{\hbar}{2M}\Box$ and  the initial
condition $\phi$ satisfies the free wave equation
\begin{equation}
\Box \phi=0
\end{equation}
then such an initial wave function gives a
 non-trivial contribution to the
average (27). We shall discuss this term later. Now, consider the covariance
\begin{equation}
\begin{array}{l}
E[((\phi_{\tau},f)-E[(\phi_{\tau},f)])((\phi_{\tau^{\prime}},f^{\prime})
-E [(\phi_{\tau^{\prime}},f^{\prime})])]=
\cr
\frac{\hbar^{2}}{2i}\Big(f,{\cal A}^{-1}
\Big(exp(-i{\cal A}|\tau-\tau^{\prime}|)-
exp(-i(\tau+\tau^{\prime}-2\tau_{0}){\cal A})\Big)f^{\prime}\Big)
\end{array}
\end{equation}
The average (27) is
\begin{equation}
\begin{array}{l}
lim_{T\rightarrow \infty}(T-\tau_{0})^{-1}
 \int_{\tau_{0}}^{T}d\tau E[((\phi_{\tau},f)-E[(\phi_{\tau},f)])
 ((\phi_{\tau},f^{\prime})
-E [(\phi_{\tau},f^{\prime})])]=
\cr
\frac{\hbar^{2}}{2i}(f,{\cal A}^{-1} f^{\prime}  )
-\frac{\hbar^{2}}{4}lim_{T\rightarrow \infty}(T-\tau_{0})^{-1}
({\cal A}^{-1}f,\Big(
exp(-2i(T-\tau_{0}){\cal A})-1\Big){\cal A}^{-1}f^{\prime})
\end{array}
\end{equation}
The limit of the second term on the r.h.s. is equal to zero under the assumption
that ${\cal A}^{-1}f $ and ${\cal A}^{-1}f^{\prime} $ exist.
So, we define ${\cal A}^{-1}$ first on a restricted set of functions
( with no support on the mass shell). However,
it is not sufficient to define quantum fields only on test functions f
with no support on the mass-shell
( $p^{2}=0 $ or more general $p^{2}=-M^{2}$
for a massive particle when ${\cal A}=-\hbar(\Box-M^{2})$). We have to specify the
correlation functions also on the mass shell.
In the distribution theory \cite{gelfand} this means an
extension of a linear functional to the whole of ${\cal S}$.
Defining the limit $T\rightarrow \infty$  is equivalent to
a particular extension .
 There is a natural
definition of this extension resulting from the
solution (30) of eq.(28) and related to the treatment of
indefinite integrals and especially oscillatory integrals.
We make the replacement ($\epsilon > 0$ )
\begin{equation}
{\cal A}\rightarrow {\cal A}-i\epsilon
\end{equation}
Then,the distribution
\begin{equation}
\Box^{-1} (x,y)=lim_{\epsilon \rightarrow 0} (2\pi)^{-d}
\int dp exp(-ip(x-y))(-p^{2}-i\epsilon)^{-1}=\triangle_{F}(x-y)
\end{equation}
coincides with Feynman$^{\prime}$s causal function which is equal to
the time-ordered  vacuum expectation value of the real scalar free field
\begin{equation}
<0|T(\phi(x)\phi(y))|0>=i\hbar\triangle_{F}(x-y)
\end{equation}
Note that if $(x-y)^{2}\neq 0$ in eq.(42) then $\triangle_{F}$
is a regular function. This property remains true for the
kernel ${\cal K}_{T}(x-y)$ of the operator
\begin{displaymath}
{\cal K}_{T}=\Big(
exp(-2i(T-\tau_{0}){\cal A})-1\Big){\cal A}^{-2}
\end{displaymath}
( eq.(40) but now with ${\cal A}=-\frac{\hbar}{2M}\Box$) . Hence, if $(x-y)^{2}\neq 0$ then the limit
$T\rightarrow \infty$ in eq.(40) holds true not only in a distributional
sense but also for the operator kernels ( in particular
 ${\cal K}_{T}(x-y)$ tends to zero for every x and y if
 $(x-y)^{2}\neq 0$). This stronger mode of convergence of the
 $\tau$-averages may be important
  for a convergence in a model with an interaction when
  there is a non-trivial renormalization and the
  distributional convergence is not sufficient.

We can generalize this result to an average of an arbitrary
 number of fields
\begin{equation}
\begin{array}{l}
<((\phi_{\tau},f_{1})-E[(\phi_{\tau},f_{1})])((\phi_{\tau},f_{2})
-E [(\phi_{\tau},f_{2})]) ....
\cr
((\phi_{\tau},f_{2n})-E[(\phi_{\tau},f_{2n})])>
=\hbar^{2n}\sum_{pairs}\prod_{(j,k)}\frac{1}{2i}(f_{j},{\cal A}^{-1}f_{k})
\end{array}
\end{equation}
where the sum is over the product of all pairs $(j,k)$ in agreement with
the Gaussian integral combinatorics ( the expectation value (44)
is equal to zero if the number of fields is odd). With ${\cal A}=-\hbar
(\Box-M^{2})$ eq.(44) can be proved by means of explicit computations
( as in eqs.(39)-(40) )  through an application of the Fourier transform.

The limit (27) exists on test functions such that ${\cal A}^{-1}f$
is well-defined i.e.$f\in Range({\cal A})$. We could instead of eq.(28)
consider the equation for ${\cal A}\phi$
\begin{displaymath}
d({\cal A}\phi  )=-i {\cal A} ({\cal A}\phi)d\tau +\sqrt{2}\hbar{\cal A}dB
\end{displaymath}
Then, we would not have the problem of an inverse in eq.(44) ( we just let
$f_{j}\rightarrow {\cal A}f_{j} $ in eq.(44) ).  We would have obtained
the correlation functions of ${\cal A}\phi=\chi$. The ambiguity in the inversion
problem arises if we wish to express $\phi$ by $\chi$ ( this problem
can be considered as another linear stochastic equation).  Such an equation in
${\cal S}^{\prime}$ poses the problem \cite{gelfand} of an extension of a linear functional
defined on a subspace of ${\cal S}$ to the whole of ${\cal S}$.

In this section we have discussed only the random field
corresponding to the quantum real scalar field. If the quantum field
has more components then we proportionally increase the number of components of the
random field (as well as the number of Brownian motions in eq.(26)).
So, for example
 we treat the complex scalar field as a real doublet $(\phi_{1},\phi_{2}) $ .
 The stochastic equation for a free electromagnetic field depends on whether
 we add the gauge fixing terms to the Lagrangian or not .
 Without any gauge fixing terms it reads
 \begin{equation}
 \partial_{\tau} A_{\mu}=\hbar i\partial^{\nu}F_{\mu\nu}d\tau+
  \sqrt{2}\hbar dB_{\mu}
  \end{equation}
  If an external current $J_{\mu}$ is added to the Lagrangian then
  the stochastic  equation takes the form
 \begin{equation}
 \partial_{\tau} A_{\mu}=i\hbar\partial^{\nu}F_{\mu\nu}d\tau-i\hbar
 J_{\mu}d\tau+
  \sqrt{2}\hbar dB_{\mu}
  \end{equation}
  We would not obtain a finite limit $T\rightarrow \infty$ of the
  correlation functions (27) of $A$ without any gauge fixing.
  However,  we may restrict ourselves to gauge invariant
  observables e.g. to $F_{\mu\nu}$ . Then ,
  we can easily rewrite eq.(46) in a gauge invariant form as an equation for
  $F_{\mu\nu}$
 \begin{displaymath}
 \partial_{\tau} F_{\alpha\mu}=i\hbar\partial^{\nu}
 (\partial_{\alpha}F_{\mu\nu} -\partial_{\mu}F_{\alpha\nu})d\tau
 -i\hbar
(\partial_{\alpha} J_{\mu}-\partial_{\mu} J_{\alpha})d\tau+
  \sqrt{2}\hbar(\partial_{\alpha} dB_{\mu}  -\partial_{\mu} dB_{\alpha})
  \end{displaymath}
We can still simplify the problem of solving this equation if
we take a divergence of both sides
\begin{equation}
 \partial_{\tau}\partial^{\mu} F_{\alpha\mu}=-i\hbar\Box
 F_{\alpha\mu}d\tau
 -i\hbar\partial^{\mu}
(\partial_{\alpha} J_{\mu}-\partial_{\mu} J_{\alpha})d\tau+
  \sqrt{2}\hbar\partial^{\mu}
  (\partial_{\alpha} dB_{\mu}  -\partial_{\mu} dB_{\alpha})
 \end{equation}
 Now, we can apply the formulas (29)-(30) where ${\cal A}=\hbar\Box$.
 The exponential of this operator and its kernel are well-known.
 We obtain a formula expressing the $\tau$-averages of $\partial^{\nu}F_{\alpha\nu} $
 by a Gaussian field $\chi_{\alpha}(x)$ with known correlation functions. The gauge problem
 appears as the non-uniqueness of the potential A solving an equation
 $\partial^{\nu}F_{\alpha\nu}=\chi_{\alpha}$ . We discuss here
 these elementary aspects of linear stochastic equations because
 they will appear in a more complex form in quantum gravity
 discussed briefly at the end of this paper.

 Eq.(45) without the noise is treated as a wave equation for the photon.
 Not surprisingly the solution is complex and
 \begin{equation}
 \overline{F_{jk}}(\tau,x)F_{jk}(\tau,x)
+ \overline{F_{0k}}(\tau,x)F_{0k}(\tau,x)
 \end{equation}
 can be interpreted as the probability density (unnormalized) of finding a photon in the space-time
 point x measured at the proper time $\tau$. Note that eq.(48) gives
 a generalization of the conventional statistical interpretation
 of the electromagnetic field when we consider
  $\tau$-independent (real) solutions ( then eq.(48) defines
 the electromagnetic energy density ).

  \section{General Lagrangians}
  We consider now an interaction among relativistic fields  $\phi_{a}$
  described by a general Lagrangian ${\cal L}(\phi_{a})$. Its
  action integral is denoted $L(\phi_{a})$ .
  The $\tau$-independent (no noise)
  solutions of the wave equations for $\phi_{a} $ should
  coincide with the classical non-linear waves.
  A proper generalization of stochastic equations
  of sec.4 reads
  \begin{equation}
  d\phi_{a}(\tau,x)=i\hbar\frac{\delta L}{\delta \phi_{a}(\tau,x)} d\tau +
 \sqrt{2}\hbar dB_{a} (\tau,x)
 \end{equation}
 where
 \begin{equation}
 E[B_{a}(\tau,x)B_{c}(s,y)]=min(\tau,s)\delta_{ac}\delta(x-y)
 \end{equation}
 We write eq.(49) in a symbolic form
 \begin{equation}
 d\phi = -i{\cal A}\phi d\tau -i gG(\phi)d\tau +\sqrt{2}\hbar dB
 \end{equation}
 Let ( as in eq.(35) ) $\Phi_{\tau}(\phi)=E[\Phi(\phi_{\tau}(\phi))]$ then it follows
 from a general theory of stochastic equations \cite{daletsky}
 that  $\Phi_{\tau}(\phi)    $ is a solution of the
 equation
\begin{equation}
\partial_{\tau}\Phi_{\tau}\equiv {\cal G}\Phi_{\tau}=\hbar^{2}Tr(D^{2}\Phi_{\tau})-i
({\cal A}\phi +gG(\phi), D\Phi_{\tau} )
\end{equation}
We can solve eq.(52) by means of an expansion in g.
\begin{equation}
\Phi_{\tau}=\sum_{n=0}^{\infty} g^{n}\Phi_{\tau}^{(n)}
\end{equation}
Then
\begin{equation}
\partial_{\tau}\Phi_{\tau}^{(n)}=\hbar^{2}Tr(D^{2}\Phi_{\tau}^{(n)})-i
({\cal A}\phi,D\Phi_{\tau}^{(n)})-i(G(\phi), D\Phi_{\tau}^{(n-1)})
\end{equation}
The solution of this equation can be expressed by means of the transition function
(33)
\begin{equation}
\Phi_{\tau}^{(n)}(\phi_{1}+i\phi_{2})=-i\int
P(\tau_{0},\phi_{1},\phi_{2};\tau,d\phi_{1}^{\prime},d\phi_{2}^{\prime})
(G(\phi^{\prime}), D\Phi_{\tau}^{(n-1)} )
\end{equation}
where $\phi^{\prime}=\phi^{\prime}_{1}+i\phi^{\prime}_{2}$.
In this way we obtain a perturbative solution of eq.(52) ( the convergence
of the series for interactions $G(\phi) $ of physical interest remains
an open problem)    .

In the Appendix we outline a proper time quantization
 of interacting quantum fields in the Fock space.
We show in the lowest order of the perturbation theory that such
a quantization leads to the same results as the stochastic quantization (49).
We supply also some non-perturbative arguments to this conjecture.
However, we think that the stochastic approach although equivalent
to the operator one avoids many difficulties related to non-commutativity.
Moreover, it  is useful for computations; in particular, for numerical
simulations ( see refs.\cite{klauder}\cite{schulke} for numerical
simulations of complex Langevin equations of the form
(49) in the context of the stochastic quantization scheme of refs. \cite{parisiwu}
\cite{parisi} ) .

\section{An average over the proper time}
When we average over the proper time $\tau$ ( as in eq.(27))   then
we obtain a linear functional $f$ on a set of functions of fields
\begin{equation}
f(\Phi)=<\Phi>
\end{equation}
We are interested in obtaining equations which could
determine the averaged values. For real processes
and real diffusion equations  the averaged value $f(\Phi)$
can be expressed by a measure called an invariant measure.
Let us recall this definition \cite{ikeda}.
 The solution    $\Phi_{\tau}(\phi)=E[\Phi(\phi_{\tau}(\phi))]$ defines
 a semigroup   $\Phi_{\tau}(\phi)\equiv (P_{\tau}\Phi)(\phi)    $ .
 We say that $\nu$ is an invariant measure if for a dense
 set of functions
 \begin{equation}
 \int d\nu(\phi) (P_{\tau}\Phi)(\phi) =
 \int d\nu(\phi)\Phi(\phi)
 \end{equation}
  So, we could say that computing the expectation values
 with respect to the noise $B$ is equivalent to a functional integral
 with respect to $\nu$.

 Differentiating eq.(57) over time  at $\tau=0 $ we obtain
 \begin{equation}
 \int d\nu(\phi) {\cal G}\Phi(\phi)\equiv \int d({\cal G}^{*}\nu)\Phi=0
 \end{equation}
 where the adjoint operator ${\cal G}^{*}$ is defined by
 the duality in the space of linear functionals.

We shall show that the averaged value (27)
is formally expressed by the Feynman integral.
In fact, the averaging over the proper time could have been considered
as a rigorous definition of the Feynman integral.
We have already calculated  the average values  as the limit
$T\rightarrow \infty$ of the correlations of the
  free fields (44). On the other hand a formal Feynman integral
 gives the same result
\begin{displaymath}
\int d\phi exp\Big(\frac{i}{\hbar}L^{free}(\phi)\Big)
\phi(x_{1}).....\phi(x_{2n})=<\phi(x_{1}).....\phi(x_{2n})>^{free}
\end{displaymath}
here
\begin{displaymath}
L^{free}(\phi)=-\frac{1}{2}\int \phi \Box \phi
\end{displaymath}
 We shall show that in each order of the perturbation expansion
 the correlation functions defined by an
 average over the proper time fulfil the same differential
 equations as the ones resulting from
  a perturbative calculation of the Feynman integral
  ( the statement has been shown by means of a formal
  argument concerning distributional limits of oscillatory integrals in refs.
 \cite{huffel}\cite{nakazato}; the method applied here has been
 introduced first in the context of the
  stochastic quantization in ref.\cite{floratos}).
We neglect problems of the ultraviolet divergencies. In order
to avoid the singularities we should  regularize the noise B.
Then, we prove that the $\tau$-averaged stochastic correlation functions
coincide with the regularized correlation functions of the conventional
QFT. In two-dimensional QFT we need no ultraviolet regularization
, hence the proof applies directly ( but because of the infrared problems
we add the mass $\Box\rightarrow \Box-M^{2}$).

As a first step in the proof we note that if
the average over the proper time exists then from eq.(52)
\begin{equation}
\begin{array}{l}
0=lim_{T\rightarrow\infty} (T-\tau_{0})^{-1}\int_{\tau_{0}}^{T} d\tau
\partial_{\tau}E[\Phi(\phi_{\tau}(\phi))]=
\cr
=lim_{T\rightarrow\infty} (T-\tau_{0})^{-1}\int_{\tau_{0}}^{T} d\tau
{\cal G}_{\phi}E[\Phi(\phi_{\tau}(\phi))]={\cal G}_{\phi}<\Phi(\phi_{\tau}(\phi))>
\end{array}
\end{equation}
If the average in the last line of eq.(59) is expressed by a (complex) measure
$\nu$ then eq.(57) defines an equation for this measure
\begin{displaymath}
{\cal G}^{*}\nu (\phi)=0
\end{displaymath}
or
\begin{equation}
D\Big( \hbar^{2}D+i{\cal A}\phi +igG(\phi)\Big)\nu =0
\end{equation}
If   ${\cal A}=-\hbar\Box $ and
\begin{equation}
G(\phi)=\hbar DV(\phi)
\end{equation}
then the solution of eq.(60) as a formal complex Feynman measure reads
\begin{displaymath}
d\nu(\phi)=d\phi exp(\frac{i}{\hbar}L(\phi))
\end{displaymath}
where
\begin{equation}
L(\phi)=\int dx{\cal L}(\phi(x))=-\int dx\Big(\frac{1}{2}\phi\Box\phi+
 gV(\phi(x))\Big)
 \end{equation}

  \section{Scalar quantum electrodynamics}

  We discuss in more detail the scalar electrodynamics
  with $(\overline{\phi}\phi)^{2}$ interaction ( this interaction is
  needed for stability in less than four dimensions and additionally for
  renormalizability in four dimensions).
  We treat a complex scalar field as a doublet $\phi_{a}$ ($a=1,2  $).
  Then eqs.(49) read  ( in the Feynman gauge; $e$ denotes the electric charge)
  \begin{equation}
  \begin{array}{l}
  d\phi_{a}=i\hbar\Box\phi_{a}d\tau+ ig\hbar (\phi_{1}^{2}+\phi_{2}^{2})\phi_{a}d\tau
  -i e\epsilon_{ab}A_{\mu}\partial_{\mu}\phi_{b}  d\tau
  -\frac{i}{\hbar} e^{2}A_{\mu}A^{\mu}d\tau +\sqrt{2}\hbar dB_{a}
  \cr
  \equiv i\hbar\Box_{A}\phi_{a}d\tau+
   ig\hbar (\phi_{1}^{2}+\phi_{2}^{2})\phi_{a}d\tau
   +\sqrt{2}\hbar dB_{a}
  \end{array}
  \end{equation}
  \begin{equation}
  dA_{\mu}=i\hbar\Box A_{\mu} d\tau+
  i\hbar e\epsilon_{ab}\phi_{b}
  \partial_{\mu}\phi_{a} d\tau +\sqrt{2}\hbar dB_{\mu}
  \end{equation}
  We write  eq.(64) in an integral form specifying the
  initial condition
   \begin{displaymath}
  \begin{array}{l}
  A_{\mu}(\tau,x)=exp(i\hbar(\tau-\tau_{0})\Box)A_{\mu} +
  ie \hbar\int_{\tau_{0}}^{\tau}exp(i\hbar(\tau-s)\Box)\epsilon_{ab}\phi_{b}
  \partial_{\mu}\phi_{a} ds +
  \cr
  +\sqrt{2}\hbar\int_{\tau_{0}}^{\tau}
  exp(i\hbar(\tau-s)\Box)dB_{\mu}(s) \equiv A_{\mu}^{cl}+A_{\mu}^{Q}
  \end{array}
  \end{displaymath}
  where we denoted by $A_{\mu}^{cl}$ the noise-independent part of $A_{\mu}$.
  Then, eq.(63) is rewritten in an integral form
  \begin{equation}
  \begin{array}{l}
  \phi(\tau,x)=exp(i\hbar(\tau-\tau_{0})\Box_{A})\phi+
  ig\hbar\int_{\tau_{0}}^{\tau}exp( i\hbar\Box_{A}(\tau-s))
 (\phi_{1}^{2}+\phi_{2}^{2})\phi(s) ds      +
\cr
   +\sqrt{2}\hbar\int_{\tau_{0}}^{\tau}exp\Big(i\hbar\Box_{A}(\tau-s)\Big)dB(s)
\end{array}
\end{equation}
It is understood that $\phi$ in this equation is a two-dimensional
vector and $A_{\mu}$ is a matrix with matrix elements $(A_{\mu})_{ab}
=\epsilon_{ab}A_{\mu}$ . If we insert 0 as  the  initial conditions
in eqs.(63)-(64)   then as follows from secs.6 and 8
after averaging over $\tau$ we obtain the standard scalar QED
\cite{schweber}. In particular, the formula (74) of the subsequent
section can be applied
for a computation of  the scattering amplitudes.
If the initial conditions $\phi$ in eq.(63) and $A^{cl}$
in eq.(64) are different from zero then our formalism goes
beyond the conventional QFT. We restrict ourselves here
to a discussion of the approximation with $e=0$ in eq.(64)
whereas in eq.(63) we neglect the noise and the terms non-linear in $\phi$.
Then, eq.(63) is just the relativistic  wave equation (1)
with the electromagnetic field A which is the sum $A^{cl}+A^{Q}$ .
In such a case the solution (65) of eq.(63) is again expressed by eq.(7)
and eq.(11). We can compute the contribution  of the quantum
electromagnetic fluctuations
upon the meson scattering processes as well as  the effect
of  these fluctuations
on the energy levels of mesonic atoms ( the Lamb shift) .
In this approximation to QFT we have a well-defined
limit $\hbar\rightarrow 0$ of QFT when the first term on the r.h.s.
of eq.(65) describes a localized trajectory of eq.(13).
The method of calculations       in the framework of
the proper time stochastic field theory may be
 considered as a rigorous version of
the argument of Welton \cite{welton}. We consider the approximation
(10)-(13). The solution of the Hamilton-Jacobi equation (4) can be
expressed by the solution of the Newton equation (14)
\begin{displaymath}
W(\tau,x)=W(\xi(\tau,x))+\int_{0}^{\tau}(\frac{M}{2}
\frac{d\xi_{\mu}}{ds} \frac{d\xi^{\mu}}{ds}  +
\frac{e}{c} A_{\mu} \frac{d\xi^{\mu}}{ds} )ds
 \end{displaymath}
 We compute perturbatively the effect of $A^{Q}$ upon
 the trajectory ( we performed similar computations in ref.\cite{haba8}) .
  We write
 \begin{displaymath}
 \xi=\xi^{(cl)}+\xi^{(Q)}=\xi^{(cl)}+\hbar \xi^{(1)}+\hbar^{2}\xi^{(2)}
 \end{displaymath}
 where
\begin{displaymath}
M\frac{d^{2}\xi^{(cl)}_{\mu}}{ds^{2}}=F^{cl}_{\mu\nu}(\xi^{(cl)})\frac{d\xi^{(cl)\nu}}{ds}
\end{displaymath}
and
\begin{equation}
M\frac{d^{2}\xi^{(Q)}_{\mu}}{ds^{2}}=
(F_{\mu\nu}(\xi^{(cl)}+\xi^{(Q)})- F^{cl}_{\mu\nu}(\xi^{(cl)}))
\frac{d\xi^{(cl) \nu}}{ds}
+F_{\mu\nu}(\xi^{cl}+\xi^{Q})\frac{d\xi^{(Q)\nu}}{ds}
\end{equation}
We can compute from eq.(66) perturbatively $\xi^{(Q)}$ in powers of
$A^{Q}$ ( with $A^{cl}=(0,0,0,-\frac{e^{2}}{r})$
we obtain Bethe$^{\prime}$s \cite{bethe}     and
Welton$^{\prime}$s $\delta({\bf x})$ correction to the potential). Then, the evolution of the relativistic wave function
in the electromagnetic field $A^{cl}+A^{Q}$ is
computed  from the formula
\begin{displaymath}
\phi_{\tau}=<<exp(i\hbar(\tau-\tau_{0})\Box_{A})\phi>>
\end{displaymath}
where $<<..>>$ means the average over the electromagnetic field
in the approximation $e=0$ in eq.(64).

  \section{Large time behavior and scattering}
If the interaction is weak at large distances then we expect
that the wave function $\phi_{\tau}$ or the corresponding
stochastic field behaves at large time as a solution of the equation
with no interaction ( eq.(28) with ${\cal A}\rightarrow {\cal A}_{0}$,
 a differential
operator with constant coefficients). Without noise such a behavior means
\begin{equation}
\phi_{\tau} \approx exp(-i\tau{\cal A}_{0})\phi_{in}
\end{equation}
when $\tau\rightarrow -\infty$  ( on a   formal level such a behavior
can be imposed if we rewrite eq.(49) in an integral form )
 and
\begin{displaymath}
\phi_{\tau} \approx exp(-i\tau{\cal A}_{0})\phi_{out}
\end{displaymath}
when $\tau\rightarrow +\infty$ .

In non-relativistic quantum mechanics if the Schr\"odinger
time evolution is determined by H
then a local distortion of the behavior (scattering) is
described by the operator
\begin{equation}
\begin{array}{l}
S=lim_{t\rightarrow \infty,t_{0}\rightarrow -\infty}
{\cal U}(t,t_{0})=
\cr
lim_{t\rightarrow \infty,t_{0}\rightarrow -\infty}exp(\frac{i}{\hbar}
H_{0}t)exp(-\frac{i}{\hbar}H(t-t_{0}))exp(-\frac{i}{\hbar}H_{0}t_{0})
\end{array}
\end{equation}
If $H-H_{0}= V$ then ${\cal U}$ satisfies the equation
( useful for perturbative calculations )
\begin{equation}
i\hbar\partial_{t}{\cal U}= V(t){\cal U}
\end{equation}
where
\begin{displaymath}
V(t)=exp(\frac{i}{\hbar}H_{0} t)V exp(-\frac{i}{\hbar}H_{0}t)
\end{displaymath}
Similar formulas hold true for a relativistic quantum mechanics
described by the Hamiltonian ${\cal A}$
\begin{equation}
\begin{array}{l}
S_{rel}=lim_{\tau\rightarrow \infty,\tau_{0}\rightarrow -\infty}
{\cal U}_{rel}(\tau,\tau_{0})=
\cr
lim_{\tau\rightarrow \infty,\tau_{0}\rightarrow -\infty}exp(i{\cal A}_{0}\tau)exp(-i{\cal A}(\tau-\tau_{0}))
exp(-i{\cal A}_{0}\tau_{0})
\end{array}
\end{equation}
If ${\cal A}-{\cal A}_{0}={\cal V}$ then ${\cal U}_{rel}$ satisfies the equation
\begin{equation}
\partial_{\tau}{\cal U}_{rel}={\cal V}(\tau){\cal U}_{rel}
\end{equation}
where
\begin{displaymath}
{\cal V}(\tau)=exp(i{\cal A}_{0}\tau)
{\cal V} exp(-i{\cal A}_{0}\tau)
\end{displaymath}
When $c\rightarrow \infty$  then ( as we have shown at the end
of sec.2 ) the evolution operator of eq.(1) tends to the evolution
operator (19) of Howland. Then, it follows from the formula
(20) ( see also \cite{howland}) that the scattering operator (70)
of the evolution equation (20) in the limit $c\rightarrow \infty$
 coincides with  S of eq.(68). Hence, $S_{rel}\rightarrow   S$.

We expect  that in the  scattering of particles
of low energy  the quantum field theory
should give the same results as the quantum mechanics .
We could establish this property in scalar QED of sec.7 when
the iterative solution A of eq.(64) starts from $A^{cl}$ such that $\Box A^{cl}=0$.
Then, $\phi_{\tau}=exp(-i\hbar\tau\Box_{A^{cl}})\phi$ plus the
noise and higher order terms . If the noise and terms of higher order
in the coupling constant are neglected then  we  obtain the
time evolution and scattering of relativistic quantum mechanics.

The quantum field theory is represented  as a non-linear stochastic
wave mechanics. The asymptotic free behavior (67) makes well sense
in such a non-linear wave mechanics. The S-matrix can be defined
as
\begin{equation}
\phi_{out}=S\phi_{in}
\end{equation}
 Its matrix elements are products of the in and out states.
 So, if the in- state  $f_{q}(x)$ satisfies ${\cal A}_{0}f_{q}=0$
( in eq.(51) we let ${\cal A}\rightarrow {\cal A}_{0}=-\hbar(\Box-M^{2}) $)
  then the matrix element is
 \begin{displaymath}
 (f_{q},\phi_{out})=lim_{T\rightarrow \infty}
 (f_{q},exp(-iT{\cal A}_{0})\phi_{out})
 \approx lim_{T\rightarrow \infty}\int_{0}^{T}
 (f_{q},\partial_{\tau}\phi_{\tau})d\tau
\end{displaymath}
 We treat the multiparticle scattering as a scattering
  of independent particles
 connected only by the common noise. Then ,  a suggested generalization  of
 the one particle matrix element to the multiparticle one could be
 \begin{equation}
 <p_{1},...,p_{n}|S|q_{1},....,q_{m}>= lim_{T\rightarrow \infty}
\frac{1}{T} \int_{0}^{T}d\tau E\Big[\prod_{j}\prod_{k}
( \overline{f_{p_{j}}},\partial_{\tau}\phi_{\tau})
(  f_{q_{k}},\partial_{\tau}\phi_{\tau})\Big]
 \end{equation}
 If we replace $\partial_{\tau}\phi_{\tau} $ by $-i{\cal A}_{0}\phi_{\tau}$
( which follows by a formal differentiation of eq.(67)), average
over $\tau$ first
( before the differentiation contained in
  ${\cal A}_{0}=\Box-M^{2}$ ) then from the stochastic field theory we obtain the standard LSZ
 formula for the scattering matrix \cite{schweber}
 \begin{equation}
 \begin{array}{l}
 <p_{1},...,p_{n}|S|q_{1},....,q_{m}>=
 \int dx_{1}.....dx_{n}dy_{1}....dy_{m}\overline{f_{p_{1}}}(x_{1})
  ....\overline{f_{p_{n}}}(x_{n})
  f_{q_{1}}(y_{1})..
  \cr
  ......f_{q_{m}}(y_{m})
 (\Box_{x_{1}}-M^{2}).....(\Box_{y_{m}}-M^{2})
  <\phi_{\tau}(x_{1}).......\phi_{\tau}(x_{n})\phi_{\tau}(y_{1})
  ...\phi_{\tau}(y_{m})>
  \end{array}
  \end{equation}

 \section{Dissipative relativistic quantum mechanics}
 Quantum theory of microscopic phenomena
 cannot be separated from a description of the
 macroscopic world of measuring devices. An interaction of
 a quantum particle with a macroscopic environment is often
 described by a random wave function or (equivalently)
 by a density matrix
 \begin{equation}
 \rho(x,y)=E[\phi(x)\overline{\phi}(y)]
 \end{equation}
 If we require that the trace and positivity of $\rho$
 are preserved by the time evolution then we obtain the Lindblad
 equation \cite{lindblad} ( Lindblad equation for the relativistic
  quantum mechanics has been discussed earlier in ref.\cite{blanchard} )
 \begin{equation}
 \partial_{\tau}\rho=-i[{\cal A},\rho]-\frac{1}{2}\sum_{k}R_{k}^{+}R_{k}\rho
          -\frac{1}{2}\sum_{k}\rho R_{k}^{+}R_{k} + \sum_{k} R_{k}\rho R_{k}^{+}
 \end{equation}
 where R describes a (phenomenological) dissipation.

We can equivalently represent the dissipative  dynamics by
the Ito stochastic wave equation (i.e. a random perturbation of eq.(1))
 which resembles eq.(26)
\begin{equation}
d\phi=-i {\cal A}\phi d\tau +i\sum_{k}R_{k}\phi d{\cal B}_{k}
-\frac{1}{2}\sum_{k}R_{k}^{+}R_{k}\phi d\tau
\end{equation}
where ${\cal B}_{k}$ are independent complex Brownian motions
\begin{displaymath}
E[{\cal B}_{k}{\cal B}_{r}]=0
\end{displaymath}
\begin{displaymath}
E[\overline{{\cal B}_{l}}(s,x) {\cal B}_{k}(\tau,y)]=\delta_{kl}min(\tau,s)
\delta(x-y)
\end{displaymath}
If the operators R are Hermitian then eq.(77) can be expressed
in a more compact form
\begin{equation}
d\phi=-i {\cal A}\phi d\tau +i\sum_{k}R_{k}\phi\circ d\tilde{B}_{k}
\end{equation}
where the circle denotes the Stratonovitch differential \cite{ikeda}
and $\tilde{B}_{k}$ are independent real Brownian motions. When we solve
eq.(77) and calculate the expectation value (75) then we can see
that the solution of eq.(77) determines a solution of the
Lindblad equation (76).     If the wave function $\phi$ has more components
then these components add discrete indices to the density matrix.
As an example , the random wave equation for the
electromagnetic field in the gauge $A_{0}=0$ reads
\begin{equation}
dA_{j}=i\hbar\partial_{0}F_{j0}d\tau
 -i\hbar\partial_{k}F_{jk}d\tau+iR_{ljk}A_{k} d{\cal B}_{l}
 -\frac{1}{2}\overline{R}_{lkj}R_{lkm}A_{m}d\tau
 \end{equation}
 We define
 \begin{displaymath}
 \rho_{jk}(x,y)=E[A_{j}(x)\overline{A}_{k}(y)]
 \end{displaymath}
 Then the Lindblad equation takes the form
 \begin{equation}
 \partial_{\tau} \rho_{jk}= i[{\cal A},\rho]_{jk}-\frac{1}{2}
 \overline{R}_{lrj}R_{lrm}\rho_{mk}
              -\frac{1}{2}\overline{R}_{lrm}R_{lrk}\rho_{jm}  +
              \overline{R}_{lkm}R_{ljr}\rho_{rm}
\end{equation}
 where
 \begin{displaymath}
 {\cal A}_{jk;lm}=\hbar\delta_{jl}\delta_{km}\Box-\hbar\delta_{km}\partial_{j}
 \partial_{l}
 \end{displaymath}
 Eq.(80) can describe a time evolution of an ensemble of photons
 ( of various polarizations corresponding to different
 indices of $\rho$) which undergoes a measurement R.
 In particular, a continuous observation of photon$^{\prime}$s
 position corresponds to $R_{ljk}=\delta_{jk}x_{j}a_{l}$
 ( where the vector ${\bf a}$ selects an orientation in space).

 Let us note that if we
add a dissipation in the form (78) to the field equation (28)
\begin{equation}
d\phi=-i {\cal A}\phi d\tau +iR\phi\circ d\tilde{B}+\sqrt{2}\hbar dB
\end{equation}
and first calculate the expectation value over B
 \begin{displaymath}
 \rho(x,y)=E\Big[E_{B}[\phi(x)]\overline{E_{B}[\phi(y)]}\Big]
 \end{displaymath}
 then we obtain again the density matrix $\rho$ satisfying eq.(76) .
 However, we  interprete eq.(81) in a different way.
 We consider eq.(81) as an equation
 for a quantum field interacting in a dissipative way with
 an environment. Then, both noises $B$ and $\tilde{B}$
 are treated on an equal footing i.e. we average
 over the noise at the end. We generalize eq.(81)
 to  the non-linear quantum field theory (51)
 allowing R to be a real non-linear function of $\phi$
\begin{equation}
d\phi=-i {\cal A}\phi d\tau -igG(\phi)d\tau
+i\gamma R(\phi)\circ d\tilde{B}+\sqrt{2}\hbar dB
\end{equation}

 Let (as in eq.(35) ) $\Phi_{\tau}(\phi)=E[\Phi(\phi_{\tau}(\phi))]$ then it
  follows from eq.(82)
 that  $\Phi_{\tau}(\phi)    $ is a solution of the
 equation
\begin{equation}
\partial_{\tau}\Phi_{\tau}\equiv \tilde{{\cal G}}\Phi_{\tau}=
\hbar^{2}Tr(D^{2}\Phi_{\tau})-\frac{\gamma^{2}}{2}Tr(R(\phi) D R(\phi) D
\Phi_{\tau})-i
({\cal A}\phi +gG(\phi), D\Phi_{\tau} )
\end{equation}
We   average again over $\tau$ and look for
 a measure $\nu$ such that an average over
$\nu$ is equivalent to the $\tau$-average.
We obtain a condition which is an analog of eq.(58)
 \begin{equation}
 \int d\nu(\phi) \tilde{{\cal G}}\Phi(\phi)\equiv \int d(\tilde{{\cal G}}^{*}\nu)\Phi=0
 \end{equation}
We rewrite eq.(84) as a differential equation for $\nu$
\begin{equation}
D\Big( \hbar^{2}D -\frac{\gamma^{2}}{2}R(\phi) R(\phi) D
+ \frac{\gamma^{2}}{2}R(\phi)DR(\phi)+i{\cal A}\phi +igG(\phi)\Big)\nu =0
\end{equation}
Let as in eq.(60)   ${\cal A}=-\hbar\Box $ and
\begin{equation}
G(\phi)=\hbar DV(\phi)
\end{equation}
We express the solution of eq.(85) as a formal complex
Feynman measure
\begin{displaymath}
d\nu(\phi)=d\phi exp(\frac{i}{\hbar}L_{\gamma}(\phi))
\end{displaymath}
where
\begin{equation}
L_{\gamma}(\phi)=\int dx{\cal L}(\phi(x))=\int dx\Big(-\frac{1}{2}\phi\Box\phi-
 gV(\phi(x))+\frac{\gamma^{2}}{2} {\cal F}(\phi)\Big)
 \end{equation}
 Then, eq.(85) for $\nu$ can be rewritten as a linear equation for ${\cal F}$
\begin{equation}
\hbar^{2} D{\cal F}=R(\phi) R(\phi)
DL_{\gamma}+ i\hbar R(\phi)DR(\phi)
\end{equation}
Its solution is a complex function of $\phi$ . The imaginary part
of ${\cal F}$ describes  a dissipation. In the lowest order in $\gamma$
we obtain
\begin{displaymath}
Im{\cal F}=\frac{1}{2\hbar}\int dx R(\phi(x))^{2}
\end{displaymath}
The Feynman integral representing the averaged correlation functions
(27) is an oscillatory integral describing the interference phenomena in
quantum mechanics. The addition of a dissipation damps the interference
leading to decoherence \cite{zurek} and a smooth equilibrium limit \cite{haba2}
represented by the complex Gibbs measure (87).

 \section{Quantum fields on a general curved manifold}
 The proper time formalism is especially useful
 if fields are to be defined on a pseudoriemannian manifold ${\cal M}$ . In such a case
 there is no candidate for a time as an evolution parameter
 ( the coordinate $x_{0}$ is only locally defined ). Moreover,
 the classical proper time (intrinsic time) is a function  of the metric. Hence,
 in quantum gravity , when the metric becomes a dynamical variable,
 then the proper time acquires the same dynamical content as the coordinate
 ${\bf x}$ in non-relativistic quantum mechanics.
We can easily generalize eq.(1) to a general pseudoriemannian
manifold ${\cal M}$ ( we set $c=1$ in this section)
\begin{equation}
i\hbar\partial_{\tau}\psi=\frac{1}{2M}g^{\mu\nu}
(-i\hbar\partial_{\mu} +A_{\mu}+\hbar g^{\alpha\beta}
\Gamma_{\mu\alpha\beta})(-i\hbar\partial_{\nu}+A_{\nu})\psi
\end{equation}
where $g^{\mu\nu}$ is the Riemannian metric and $\Gamma$ is the
Christoffel symbol of the Levi-Civita connection .
Assume that $W_{\tau}$ is a solution of the Hamilton-Jacobi equation
\begin{equation}
\partial_{\tau} W_{\tau}+
\frac{1}{2M}g^{\mu\nu}(\partial_{\mu} W_{\tau} +A_{\mu})
(\partial_{\nu}W_{\tau}+A_{\nu}) =0
\end{equation}
with the initial condition W.
Let us consider eq.(89) with the initial condition $\phi=exp(\frac{i}{\hbar} W)
\Phi$ . Then, $\phi_{\tau}   $ is a solution of eq.(89) if and only if
$\Phi_{\tau}$ is the solution of the equation  ( the Lorentz gauge for $A_{\mu}$
is assumed )
\begin{equation}
\begin{array}{l}
i\hbar\partial_{\tau}\Phi=-\frac{\hbar^{2}}{2M}\Box_{g}\Phi-
  \frac{i\hbar}{M}g^{\mu\nu}(\partial_{\mu}W_{\tau}
 +A_{\mu})\partial_{\nu}\Phi_{\tau} -\frac{i\hbar}{2M}\Box_{g}\Phi_{\tau}
\end{array}
\end{equation}
where $\Box_{g}$ is the wave operator
on the pseudoriemannian manifold ${\cal M}$
\begin{displaymath}
\Box_{g}=g^{\mu\nu}\partial_{\mu}\partial_{\nu}+\frac{1}{2}g^{\nu\rho}
\Gamma^{\mu}_{\nu\rho}\partial_{\mu}
\end{displaymath}
In the formal limit $\hbar\rightarrow 0$ of eq.(91) we obtain
$\phi_{\tau}(x)\approx\Phi(\xi(\tau))$ where
 $\xi$ is the solution of the equation   ($0\leq s \leq \tau $)
\begin{equation}
\frac{d\xi^{\mu}}{ds}=-\frac{1}{M}g^{\mu\nu}(\xi(s))
\Big(\partial_{\nu}W(\tau-s,\xi(s))+A_{\nu}(\xi(s))\Big)
\end{equation}
Differentiating eq.(92) once more and using the Hamilton-Jacobi equation
(90) we obtain the  equation
\begin{equation}
\frac{d^{2}\xi^{\mu}}{ds^{2}}+\Gamma^{\mu}_{\nu\rho}\frac{d\xi^{\nu}}{ds}
\frac{d\xi^{\rho}}{ds}=\frac{1}{M}F^{\mu\nu}(\xi(s))\frac{d\xi_{\nu}}{ds}
\end{equation}
 Eq.(93) shows  that the correct classical   limit
( as a geodesic equation ) of a motion of the quantum  particle on a general
pseudoriemannian manifold results if and only if $\tau$ is the proper time.
The interpretation of $\tau $ as the classical proper time remains
true not only in the leading order in $\hbar$ but also in all subsequent
terms because the leading order determines the subsequent interpretation of
$\tau$.

We can express the quantum corrections in terms of the Brownian motion.
Let us define the complex vierbeins $v_{\mu a}$
\begin{equation}
v_{\mu a}v_{\nu a}=ig_{\mu\nu}
\end{equation}

Then, we consider a complex Markov process q as a solution of the set of
covariant equations  ( for the case of the Riemannian manifold and
imaginary time see
 \cite{ikeda} or \cite{cdewitt}; the real time is discussed in \cite{haba3},
 it can be considered as a complexification of the diffusion  )
\begin{equation}
dq_{s}^{\mu}=-\frac{1}{M}g^{\mu\nu}(q_{s})\Big(
\partial_{\nu} W(\tau-s,q_{s})+A_{\nu}(q_{s})\Big) ds
+\sigma v^{\mu}_{a}(q_{s})\circ db^{a}
\end{equation}
\begin{equation}
dv^{\mu}_{a}+\Gamma^{\mu}_{\nu\rho}v^{\nu}_{a}\circ dq^{\rho}=0
\end{equation}
Then, the solution of eq.(91) with the initial condition $\psi=
exp(\frac{i}{\hbar} W)\Phi$ reads
\begin{equation}
\psi_{\tau}(x)=exp(\frac{i}{\hbar}W_{\tau}(x))
E\Big[exp\Big(-\frac{1}{2M}\int_{0}^{\tau}\Box_{g} W(\tau-s,q(s))ds\Big)
\Phi(q(\tau))\Big]
\end{equation}
Eq.(97) can be applied for a semiclassical expansion in powers of $\hbar$.

We can continue now (as in the earlier sections) with
the quantization of the wave equation (89). So, for the free field
the stochastic equation coincides  with eq.(28) where
\begin{equation}
{\cal A}=-\hbar\Box_{g}
\end{equation}
In the computation of an average over the proper time (27)
we need to define $\Box_{g}^{-1}$. At finite T we obtain a linear functional
on a  subset of test-functions ${\cal S}({\cal M})$. Taking the limit
$T\rightarrow \infty$ is equivalent to a definition of
 an extension of this linear
functional.
 The definition of the inverse
is not unique but it is restricted by the requirements that
the quantum field theory resulting from  $\Box_{g}^{-1}$
is causal and that the quantum fields
should be defined in a Hilbert space ( with a non-negative
scalar product).
It remains unclear whether such a definition of
$\Box_{g}^{-1}$ exists on an arbitrary manifold.
This can be achieved
on a globally hyperbolic manifold  \cite{birrel}\cite{ellis}
\cite{kay}\cite{dewitt2} . In
such a case there exists a complete set of solutions of the wave equation
\begin{equation}
\Box_{g}u_{j}=0
\end{equation}
Then, we can define
\begin{equation}
\triangle_{g}^{(+)}(x,y)=\sum_{j}\overline{u}_{j}(x)u_{j}(y)
\end{equation}
Eq.(100) determines a non-negative bilinear form. We can define
by means of the standard Fock space methods the quantum
annihilation $\phi^{(+)}(x)$
( as $\phi^{(+)}=\sum a_{j}u_{j}$) and creation $\phi^{(-)}(x)$ parts
of the field operator $\phi=\phi^{(+)}+\phi^{(-)}$. Now,
on a globally hyperbolic manifold
there exists a  choice of the $x_{0}$-coordinate \cite{kay}
such that 
\begin{equation}
\Box_{g}^{-1}(x,y)\equiv \triangle_{F}(x,y)=\theta(x_{0}-y_{0})
\triangle_{g}^{(+)}(x,y)
                   +\theta(y_{0}-x_{0})\triangle_{g}^{(+)}(y,x)
 \end{equation}
 is independent of the choice of coordinates.
Eq.(101) defines the Feynman propagator
on a globally hyperbolic manifold . The definition (101) coincides
with the one resulting from the $\epsilon$-prescription
\cite{dewitt}  ($\epsilon >0$)
\begin{equation}
\Box_{g}^{-1}= lim_{\epsilon\rightarrow 0}\int_{0}^{\infty}
i ds exp(-is\Box_{g}-\epsilon  s)
\end{equation}
The interacting field is defined by the stochastic equation (49)
with ${\cal A}=-\hbar\Box_{g}$.
After the propagator is defined  there is no further
difficulty (up to the ultraviolet problems) in defining perturbatively
an interacting field by means of an iterative solution of the stochastic equation (49).

A really ambitious program concerns a quantization of the metric g.
The proper time approach supplies a useful method at the
problem where other quantization methods
encounter insurmountable difficulties. The Feynman integral is a formal
tool . It can be defined either in the imaginary time or for complex coordinates
\cite{haba}. However, these methods fail for Einstein gravity which
has the action unbounded from below. The method
of Euclidean stochastic quantization  \cite{halpern}\cite{haba4}
works well . However, it remains unclear
whether the model can be continued back to the real time
( pseudoriemannian manifold) . An application of the
stochastic quantization \cite{parisiwu} with a complex action and real time to the Einstein
gravity has been suggested by Rumpf \cite{rumpf}. We interprete
the fictitious time of Parisi and Wu as the proper time with a
physical meaning. Such an interpretation should have experimental
consequences  for a time evolution of the graviton wave function
( in ref. \cite{rabin} an  experiment showing the
 interference in time is suggested).

We need now the Brownian motion B which
 is a symmetric matrix whose entries
 constitute independent Brownian motions
 \begin{displaymath}
 E[B_{ac}(s,x)B_{fd}(\tau,y)]=(\delta_{af}\delta_{cd} +
 \delta_{ad}\delta_{cf})min(\tau,s)\delta (x-y)
 \end{displaymath}
Then, we introduce the infinite dimensional pseudoriemannian manifold
 ${\cal G}({\cal M})$
of metric tensors $g$ on ${\cal M}$ with
 DeWitt$^{\prime}$s supermetric \cite{dewitt3} on this manifold
 \begin{equation}
 G^{\alpha\beta;\mu\nu}=(det(-g) )^{- \frac{1}{2}}(g^{\mu\alpha}g^{\nu\beta}
 +g^{\mu\beta}g^{\nu\alpha} )
 \end{equation}
 We need a basis of complex frames on ${\cal G}({\cal M})$ (tetrads) fulfilling
 the relation
 \begin{displaymath}
 G^{\mu\nu;\alpha\beta}={\cal E}^{\mu\nu;ac}{\cal E}^{\alpha\beta;ac}
 \end{displaymath}

If L is the action for the metric and the matter fields then the stochastic
equation for g according to the prescriptions (49) and (95)-(96) reads
( see refs. \cite{cdewitt}\cite{halpern}\cite{haba4}  for the case of infinite dimensional
Riemannian manifolds )
\begin{equation}
dg^{\mu\nu}=i\hbar\frac{\delta L}{\delta g_{\mu\nu}}d\tau+\hbar
{\cal E}^{\mu \nu;ac}\circ dB_{ac}
\end{equation}
\begin{equation}
d{\cal E}^{\mu\nu;ac}+
\Gamma^{(\mu\nu)}_{(\alpha\beta)(\gamma\rho)}
{\cal E}^{\gamma\rho;ac}\circ dg^{\alpha\beta} =0
\end{equation}
Eqs.(104)-(105) need a regularization of the Brownian motion B for
a proper interpretation of the Christoffel symbol
appearing in eq.(105) ( see \cite{halpern}).
Note that if L is a sum of the Einstein action and the action for the
 matter fields then eq.(104) reads
\begin{equation}
dg^{\mu\nu}=i\hbar(det(-g) )^{ \frac{1}{2}}(R^{\mu\nu}-\frac{1}{2}g^{\mu\nu}R)d\tau
-i\hbar\kappa (det(-g) )^{ \frac{1}{2}}T^{\mu\nu}d\tau+
\hbar{\cal E}^{\mu\nu;ac}\circ dB_{ac}
\end{equation}
where $\kappa$ is proportional to the gravitational constant and
 $T_{\mu\nu}$ is the energy-momentum tensor for the matter fields.
If we start from a linearized form of eq.(106) then
the first difficulty which we encounter involves the zero modes of the linear
part which cause a trouble with the large time limit. We can
deal with the problem of zero modes either by means of the stochastic
gauge fixing \cite{zwan} or restricting ourselves to gauge invariant observables.
We would like to outline here the method introduced for the electromagnetic
field at eq.(47)
( the Einstein tensor $G^{\mu\nu}=
(det(-g) )^{ \frac{1}{2}}(R^{\mu\nu}-\frac{1}{2}g^{\mu\nu}R)$
 is an analog of $F_{\mu\nu}$ , it has a gauge invariance resembling the
 one of non-Abelian gauge fields \cite{dewitt3}). So, for pure gravity
 ( $T_{\mu\nu}=0$) we have

\begin{equation}
d( (det(-g) )^{ \frac{1}{2}}(R^{\mu\nu}-\frac{1}{2}g^{\mu\nu}R))=
\frac{\delta L}{\delta g_{\alpha\beta}
\delta g_{\mu\nu}} \circ dg_{\alpha\beta}
\end{equation}
or more explicitly
\begin{equation}
\begin{array}{l}
d\Big( (det(-g) )^{ \frac{1}{2}}(R^{\mu\nu}-\frac{1}{2}g^{\mu\nu}R)\Big)=
\cr
i\hbar\frac{\delta L}{\delta g_{\alpha\beta}
\delta g_{\mu\nu}}
\Big( (det(-g) )^{\frac{1}{2}}(R^{\alpha\beta}-\frac{1}{2}g^{\alpha\beta}R))d\tau
     -i  {\cal E}^{\alpha\beta;ac}\circ dB_{ac}\Big)
\end{array}
\end{equation}
Let $g_{\mu\nu}=g_{\mu\nu}^{(0)}+h_{\mu\nu}     $ where  $g_{\mu\nu}^{(0)}   $
is the constant metric tensor of the  Minkowski space-time.
Denote  by G(h) the  part of the Einstein tensor linear in h. Then eq.(108)
reads
\begin{equation}
dG^{\mu\nu}(h)=2i\hbar\Box  G^{\mu\nu}(h)d\tau+ p(G(h))^{\mu\nu}d\tau +
{\cal C}^{\mu\nu;ac}(G(h))\circ dB_{ac}
\end{equation}
where $p$ and ${\cal C}$ are non-linear in $G(h)$. In eq.(109) we have
inverted the linear relation $h\rightarrow G(h)$ i.e. we have
chosen a particular solution $h_{\mu\nu}$ of the equation $G_{\mu\nu}(h)=
G_{\mu\nu}$.
We can solve eq.(109)  iteratively ( we assume a
coordinate invariant regularization of the ultraviolet divergencies \cite{halpern}) .
Then , the proper time average (27) of $G_{\mu\nu}(h)$
 exists in a perturbation
expansion with
$\Box^{-1}$ defined by $\triangle_{F}(x,y)$.
 We compute in this way
 the time-averaged correlation functions of   G (h).
 These correlation functions determine a random variable $\chi_{\mu\nu}$.
 Then, we can  obtain the metric tensor as a particular solution of the linear equation
 $G_{\mu\nu}(h)=\chi_{\mu\nu}$ .  Subsequently,
  we can express $g_{\mu\nu}$ in terms
 of $\chi_{\mu\nu}$ . We expect that in spite of the arbitrariness
 (gauge dependence) of the relation between G(h) and h the coordinate
 independent variables e.g. $R_{\mu\nu\rho\gamma}(g) R^{\mu\nu\rho\gamma}(g) $
 ( where $g_{\mu\nu}=g_{\mu\nu}^{(0)}+h_{\mu\nu}$ and
 R is the Riemannian curvature ) will not depend on the choice
 of the gauge ( as in the case of the stochastic gauge fixing).
In general, we could start from a background metric $g_{\mu\nu}^{(0)}$
 instead of the Minkowski one.  In such a case in addition to the processes
 of graviton creation and annihilation ( and mutual scattering) we could
 describe the time evolution and localization of a single graviton
 in interaction with other fields.
 \section{Summary}
 We have discussed a method of quantization which can be
 considered as a version of the Feynman integral.
 The method is related to the stochastic quantization of
 Parisi and Wu \cite{parisiwu}.
 We considered  the Bose fields only. If we are  interested in correlation functions
 of Fermi fields then we need stochastic equations with Grassmann
 variables.
 Our main emphasis was on the proper time interpretation of
 the "fictitious time parameter".  The proper time interpretation
 supplies a relativistically covariant quantization.
 It can give a meaning to the probability of detecting a particle
 inside a bounded region (sec.7) . It  is rather difficult
to formulate such questions  in the framework of the conventional QFT. Moreover,
 a proper time interpretation can have experimental consequences
 for time measurements \cite{rabin}.
 For quantum field theory on a manifold and quantum gravity it may be
 valuable to treat the background metric and gravitons at the same footing. The conventional Euclidean
 quantization of Einstein gravity is impossible because  the
  action is unbounded from below. The stochastic Euclidean quantization of gravity
 is feasible \cite{halpern}\cite{haba4}. However, in such an
 approach it remains unclear how to continue to the real physical
 time. The direct  quantization in the real time avoids these
 problems.
 We have clarified some mathematical aspects of
 the proper time stochastic quantization scheme.
  We have shown the
 convergence of the averaged values. The convergence
 rate of the average over  the time interval $[0,T]$ is usually proportional to $T^{-1}$. With purely Hamiltonian dynamics it
 cannot be as fast as in the Euclidean version of the stochastic quantization
  where the convergence can be exponential. Nevertheless, in view
 of the difficulties of the conventional oscillatory Feynman integral
 the stochastic version can be a useful tool for numerical
 calculations (  stochastic simulations  of complex stochastic equations
 are discussed in \cite{klauder}\cite{schulke}\cite{haba3}).
 In the dissipative quantum field theory of sec.9 the convergence
 of the averaged correlation functions can again be exponential
 as in the Euclidean framework.

\section{Appendix}
We outline in this section an operator quantization in the proper time.
The equation of motion reads
$$
  \partial_{\tau}\phi(\tau,x)=i\hbar\frac{\delta L}{\delta \phi(\tau,x)}
\eqno(A.1)
$$
 We are looking for a solution of this equation with an initial condition
 at $-\infty $
 $$
 lim_{\tau\rightarrow -\infty}\phi_{\tau}(x)=\phi_{in}(x)
\eqno(A.2)
$$
 where $(\Box-M^{2}) \phi_{in}$=0 is the quantum scalar free field of mass $M$.
 We expect that as $\tau\rightarrow \infty$
 $$
  \phi_{\tau}(x)\rightarrow \phi_{out}(x)
 \eqno(A.3) $$
 If this asymptotic behavior holds true then we can define
 $$
 \phi_{out}=S^{-1} \phi_{in} S
 \eqno(A.4)$$

 We can develop a perturbative theory now. Assume the Lagrangian (62),
 then we can rewrite   eq.(A.1) as an integral equation
$$
\phi_{\tau}(x)=\phi_{in}(x)-ig\hbar\int_{-\infty}^{\tau}
exp\Big(-i\hbar(\Box-M^{2})(\tau-s)\Big)V(\phi_{s})ds
\eqno(A.5)$$
We are interested in computing the time-ordered products
of quantum fields.
We first solve eq.(A.5) perturbatively   ( fixing the initial
condition at $\tau_{0}\rightarrow -\infty $)
$$
\begin{array}{l}
\phi_{\tau}=\phi_{in}-ig\hbar\int_{\tau_{0}}^{\tau}
exp\Big(-i\hbar(\Box-M^{2}) (\tau-s)\Big)V(\phi_{in})ds
\cr
-(ig\hbar)^{2}\int_{\tau_{0}}^{\tau}
exp\Big(-i\hbar(\tau-s)(\Box-M^{2})\Big)V(\phi_{in})ds
\cr
\int_{\tau_{0}}^{s}
exp\Big(-i\hbar(\Box-M^{2})(s-s^{\prime})\Big)V(\phi_{in})ds^{\prime}+....
\end{array}
\eqno(A.6) $$
We assume that V is defined by the Wick normal product. Then , the Wick
theorem allows us to calculate explicitly $<O|T(\phi_{\tau}(x)
\phi_{\tau^{\prime}}(x^{\prime})) |0>$ where the T-ordering is understood
in $x_{0}$ rather than in $\tau$. Subsequently, the average over the
proper time can be calculated. We obtain a closed formula in terms
of the Green$^{\prime}$s functions of $exp(-is(\Box-M^{2}))$ and $(\Box-M^{2})^{-1}$.
We must define $\Box^{-1}(x,y)$ as $\triangle_{F}(x,y) $ if
the time-ordered correlation functions are to agree with the conventional ones.
We have checked
the agreement with the conventional QFT only at the lowest order of g.
We have obtained  the conventional QFT  because of the
asymptotic condition (A.2). Eq.(A.1) at $g=0$ has more solutions
which in general are $\tau$-dependent.  Starting with such solutions
we would go beyond the conventional QFT.

Another ( implicit) argument concerning the relation
 to the conventional QFT at each order of g comes from the identity
 $$
 \begin{array}{l}
0 = lim_{R\rightarrow\infty}\frac{1}{R}\int_{0}^{R}   d\tau
<T\Big(\partial_{\tau}(\phi_{\tau}(x_{1})......\phi_{\tau}(x_{n}))\Big)>  =
 lim_{R\rightarrow\infty}\frac{1}{R}\int_{0}^{R} d\tau
 \cr
\sum_{k=1}^{n}<T\Big(\phi_{\tau}(x_{1})...\Big((\Box-M^{2}) \phi_{\tau}(x_{k})
-gV^{\prime}(\phi_{\tau}(x_{k})\Big)
...\phi_{\tau}(x_{n})\Big)>
\end{array}
\eqno(A.7) $$
We can use this equation in order to prove that if at $g=0$ the
time-ordered correlation functions
 coincide with the conventional free ones  then at each order of $g$
 the proper time average coincides with the one resulting
 from the solution of the operator equation
 \begin{displaymath}
( \Box -M^{2})\phi(x)
=gV^{\prime}(\phi(x)  )
\end{displaymath}


\begin{thebibliography}{99}
\bibitem{hepp}
K. Hepp, Comm.Math.Phys.{\bf 35},265(1974)
\bibitem{stu}
E.C.G. Stueckelberg, Helv.Phys.Acta.{\bf 14},372(1941),{\bf 14},588(1941)
\bibitem{hor}
L.P. Horwitz and C. Piron, Helv.Phys.Acta {\bf 46},316(1973)
\bibitem{feyn}
R.P.Feynman,Phys.Rev.{\bf 80},440(1950)
\bibitem{col}
R.E.Collins ,Lett.Nuovo Cim.{\bf 18},581(1977)
\bibitem{landau}
L.D. Landau and E.M. Lifshits, Mechanics, Pergamon Press,Oxford,1959
\bibitem{landautp}
L.D. Landau and E. M. Lifshits, Field Theory, Pergamon Press, Oxford,1959
\bibitem{goldstein}
H.Goldstein, Classical Mechanics , Addison-Wesley,1980
\bibitem{dirac}
P.A.M. Dirac, Lectures on Quantum Mechanics, Belfer Graduate School of
Sciences, Yeshiva University,1964
\newline
A.J. Hansson, T.Regge and C.Teitelboim, Constrained Hamiltonian Systems,
Accademia dei Lincei,Roma,1976
 \bibitem{haba}
 Z.Haba, Journ.Math.Phys.{\bf 35},6344(1994)
 \newline
 Lett.Math.Phys.{\bf 37},223(1996)
 \bibitem{haba3}
 Z. Haba, Feynman Integral and Random Dynamics in Quantum Physics,
 Kluwer Academic Publishers,Dordrecht,1999
 \bibitem{ikeda}
N. Ikeda and S. Watanabe, Stochastic Differential Equations and
Diffusion Processes, North Holland, Amsterdam,1981

 \bibitem{horob}
 L.P. Horwitz and F.C. Rotbart, Phys.Rev. {\bf D24},2127(1981)
 \bibitem{howland}
 J.S. Howland, Math.Ann.{\bf 207},315(1974)
\bibitem{gelfand}
I.M. Gelfand and N.Y. Vilenkin, Generalized Functions, Vol.1,Academic Press,1964
\bibitem{daletsky}
Y. Daletsky, Russian Math.Surv.  {\bf 22},No.4,p.1(1967)
\bibitem{klauder}
D.J.E. Callaway, F. Cooper, J.R. Klauder  and H.A. Rose, Nucl.Phys.
{\bf 262},19(1985)
 \bibitem{schulke}
 K. Okano, L. Sch\"ulke and B. Zheng,
 \newline
  Progr.Theor.Phys.Suppl.{\bf 111},313(1993)
\bibitem{parisiwu}
G. Parisi and Yong-Shi Wu ,Sci.Sin.{\bf 24},483(1981)
\bibitem{parisi}
G. Parisi, Phys.Lett.{\bf 131B},393(1983)
 \bibitem{huffel}
 H. H\"uffel and H. Rumpf, Phys.Lett.{\bf 148B},104(1984)
\bibitem{nakazato}
H. Nakazato and Y. Yamanaka,Phys.Rev.{\bf D34},492(1986)
\bibitem{floratos}
E. Floratos and J. Iliopoulos, Nucl.Phys.{\bf B214},392(1983)
\bibitem{schweber}
S.S. Schweber, An Introduction to Relativistic Quantum Field Theory,
Row and Peterson,Evanston,1961
\bibitem{welton}
T.A. Welton, Phys.Rev.{\bf 74},1157(1948)
\bibitem{haba8}
Z. Haba, Journ.Math.Phys.{\bf 39},1766(1998)
\bibitem{bethe}
H. A. Bethe, Phys.Rev.{\bf 72},339(1947)
 \bibitem{lindblad}
G. Lindblad ,Comm.Math.Phys.  {\bf 48},119(1976)
 \bibitem{blanchard}
 Ph. Blanchard and A.Z. Jadczyk ,Found.Phys.{\bf 26},1669(1996)
 \bibitem{zurek}
 W.H. Zurek,Physics Today,{\bf 44},No.10,p.36(1991)
 \bibitem{haba2}
 Z.Haba, Phys.Rev.{\bf A57},4034(1998)
  \bibitem{birrel}
 N.D. Birrell and P.C.W, Davis , Quantum Fields in Curved Space,
 Cambridge Univ.Press,1982
 \bibitem{ellis}
 S.W. Hawking and G.F.R. Ellis, The Large Scale Structure of The Space-Time,
 Cambridge Univ.Press,1973
 \bibitem{kay}
 B.S. Kay, Comm.Math.Phys.{\bf 71},29(1980)
 \bibitem{dewitt2}
 B.S.DeWitt, Phys.Rep.{\bf 19C},295(1975)
 \bibitem{dewitt}
 B. DeWitt, Quantum gravity. A new synthesis,in General Relativity.An Einstein Centenary Survey,ed.
 S.W. Hawking and W. Israel, Cambridge,1979
 \bibitem{cdewitt}
 K. D. Elworthy, Stochastic Differential Equations on Manfolds,
 Cambridge Univ.Press,1982
 \bibitem{rumpf}
 H.Rumpf,Phys.Rev.{\bf D33},942(1986),
  Progr.Theor.Phys.Suppl.{\bf 111},63(1993)
 \bibitem{halpern}
 M. Halpern, in Probabilistic Methods in Quantum Field Theory
 \newline
  and Quantum Gravity,
  P.H. Damgaard, H.H\"uffel and
 \newline
  A. Rosenblum, eds.,Plenum Press,
 New York,1990
 \bibitem{haba4}
 Z. Haba, Journ.Phys. {\bf A18},1641(1985)
\bibitem{rabin}
 L.P. Horwitz and Y. Rabin , Lett.Nuovo Cim.{\bf 17},501(1976)
 \bibitem{dewitt3}
 B. S. DeWitt ,Dynamical Theory  of Groups and Fields, Gordon and Breach,1965
 \bibitem{zwan}
 D. Zwanziger, Nucl.Phys.{\bf B192},259(1981)
 \end{thebibliography}
 \end{document}